\documentclass{article}
\usepackage{PRIMEarxiv}
\usepackage[T1]{fontenc}
\usepackage[utf8]{inputenc}
\usepackage{newtxtext,newtxmath}
\usepackage{microtype}

\usepackage{amsmath,amssymb,amsfonts,mathtools}
\usepackage{booktabs,tabularx,array}
\usepackage{url}
\usepackage[numbers,sort&compress]{natbib}
\usepackage{float}
\usepackage{graphicx}
\usepackage[hidelinks]{hyperref}
\usepackage{fancyhdr}
\allowdisplaybreaks[2]
\newcommand{\ii}{\mathrm{i}}
\newcommand{\Om}{\omega_{D}}
\newcommand{\NI}{\mathcal{N}}
\newcommand{\gamrat}{\gamma_{21}}

\renewcommand{\arraystretch}{1.13}

\title{Reduced-Order Quadratic Representation of Higher-Order Nonlinear Dynamical Systems for Phononic Frequency Comb Generation}
\author{%
 Madhurendra Mishra$^{1}$\thanks{ORCID: 0009-0003-0080-6186}\\
 $^{1}$Department of Physics, Sri Guru Tegh Bahadur Khalsa College,\\
 University of Delhi, Arts Faculty Rd, University Enclave,\\
 New Delhi, Delhi 110007, India\\
 \texttt{madhurendramishra24@gmail.com}
 \And
 Adarsh Ganesan$^{2}$\thanks{Corresponding author. ORCID: 0000-0002-5107-8452}\\
 $^{2}$Department of Electrical and Electronics Engineering,\\
 Birla Institute of Technology and Science Pilani, Dubai Campus,\\
 Dubai International Academic City, Dubai--345055, UAE\\
 \texttt{adarsh@dubai.bits-pilani.ac.in}
}
\date{}

\begin{document}
\maketitle

\begin{abstract}
Phononic frequency combs arise from nonlinear interactions between vibrational modes, and their formation can be affected by nonlinearities beyond quadratic coupling. We consider a driven two-mode system near a $2{:}1$ internal resonance and examine polynomial interactions through sixth order. Through analytical derivation, we found that any higher-order nonlinear dynamical system can be expressed as a reduced quadratic framework. In the near-resonant reduced model, it has been shown that the odd-order nonlinear terms modify the effective modal frequencies, while the even-order terms modify the quadratic coupling coefficients. We assessed the accuracy of the reduced-order quadratic representation through numerical simulations using an illustrative set of parameters representative of a MoS$_2$ resonator. Numerical onset boundaries identify force--frequency regions supporting frequency combs, while Fourier spectra reveal phononic comb sidebands around the driven and subharmonic frequencies. The Long and Short equations reproduce several common trends, although differences remain in the spectral distribution and in some numerically sensitive onset estimates. These results show how higher-order nonlinearities affect comb formation and help define the useful range of the reduced description.
\end{abstract}

\section{Introduction}
\label{sec:introduction}
When nonlinear interactions link the motion of two vibrational modes, a resonator driven at a single frequency can develop additional, regularly spaced spectral components. Such phononic frequency combs (PFCs) bring the concept of frequency comb familiar from optics to mechanical systems, where the spectrum reflects energy exchange between coupled modes rather than optical four-wave mixing \cite{Cundiff2003,Kippenberg2011}. Following an early theoretical proposal \cite{Cao2014}, the generation of phononic combs was experimentally demonstrated in a piezoelectric microresonator \cite{Ganesan2017} and examined through several modal coupling pathways \cite{Ganesan2018,Ganesan2018b,Ganesan2018c}. Subsequent studies found related responses in silicon structures, nitride membranes and strings, quartz resonators, two-dimensional materials, and other nonlinear mechanical platforms \cite{Czaplewski2018,Park2019,Singh2020,Goryachev2020,Kubena2020,Chiout2021,Yousuf2023,Keskekler2022,Ochs2022,Jong2023,Yang2021,Gobat2023,Sun2023}. These observations show that comb formation is not restricted to a particular material or actuation mechanism. What matters for theoretical interpretation is how the drive, modal detuning, damping, and nonlinear coupling determine the existence regime of frequency combs.

A central step towards that understanding was the analytical existence condition obtained by Qi et al.\ for two interacting phonon modes \cite{Qi2020}. Their treatment established how quadratic coupling constrains the detuning and amplitude required for PFC generation. However, real mechanical structures need not be described by quadratic interactions alone. Cubic and higher order nonlinearities can exist. The influence of Duffing nonlinearity on the existence conditions of a two-mode coupled model was recently examined in \cite{mishra2026icnda}. However, higher-order nonlinearities beyond cubic can also occur in micro- and nanomechanical devices \cite{Nichol2009,Kacem2015,Kumar2017,Shaw2017,Samanta2018,Nathamgari2019,Huang2019,Miller2021,Kaisar2022,Kanj2023}. A direct treatment of every nonlinear monomial becomes unwieldy as the order increases. It is therefore useful to determine which higher-order terms affect the resonant dynamics and whether their effects can be incorporated into the established quadratic-coupling description.

Here we develop a reduced two-mode model for a higher-order nonlinear system, and work out its coefficients explicitly through sixth order. Near the $2{:}1$ internal resonance, the amplitude-based resonant reduction maps the odd-order nonlinearities onto corrections to the modal natural frequencies and the even-order nonlinearities onto corrections to the effective quadratic coupling coefficients. The resulting equations preserve the quadratic form needed for the existence-boundary analysis, while retaining the amplitude dependence induced by higher-order interactions. We then assess this approximation numerically by comparing the unreduced and reduced displacement equations for cumulative nonlinear orders three to six, using a MoS$_2$-informed, explicitly identified illustrative parameter set. The comparison focuses on the numerical onset boundaries and frequency spectra of both modes, while emphasizing that the approximation is valid only within the near-resonant regime.

\section{Mathematical Modelling}\label{sec:theory}

\subsection{Coupled-mode equations}\label{subsec:full}
We consider two damped mechanical modes with natural frequencies $\omega_1$ and $\omega_2\simeq\omega_1/2$. The first mode is driven at angular frequency $\omega_D$, while the second is excited through nonlinear coupling. Their equations are
\begin{align}
\ddot{x}_1+2\gamma_1\dot{x}_1+\omega_1^2x_1+\NI_N(x_1,x_2)&=F\cos(\Om t),\label{eq:full1}\\
\ddot{x}_2+2\gamma_2\dot{x}_2+\omega_2^2x_2+\NI_N(x_1,x_2)&=0.\label{eq:full2}
\end{align}
Here $x_1$ and $x_2$ are the modal displacements, $\gamma_1$ and $\gamma_2$ are the damping rates, and $F$ is the drive amplitude. We write the nonlinear interaction, including terms up to order $N$, as
\begin{equation}\label{eq:poly}
\NI_N(x_1,x_2)=\sum_{n=2}^{N}\sum_{j=1}^{n+1}\alpha_{n,j}x_1^{\,n-j+1}x_2^{\,j-1}.
\end{equation}
The coefficients $\alpha_{n,j}$ describe the nonlinear contributions of the two modes. For the third- through sixth-order systems studied here, the interaction is cumulative, $\NI_N=\sum_{n=2}^{N}N_n$, with
\begin{align}
N_2={}&\alpha_{2,1}x_1^2+\alpha_{2,2}x_1x_2+\alpha_{2,3}x_2^2,\label{eq:poly2}\\
N_3={}&\alpha_{3,1}x_1^3+\alpha_{3,2}x_1^2x_2
 +\alpha_{3,3}x_1x_2^2+\alpha_{3,4}x_2^3,\label{eq:poly3}\\
N_4={}&\alpha_{4,1}x_1^4+\alpha_{4,2}x_1^3x_2
 +\alpha_{4,3}x_1^2x_2^2\nonumber\\
 &+\alpha_{4,4}x_1x_2^3+\alpha_{4,5}x_2^4,\label{eq:poly4}\\
N_5={}&\alpha_{5,1}x_1^5+\alpha_{5,2}x_1^4x_2
 +\alpha_{5,3}x_1^3x_2^2\nonumber\\
 &+\alpha_{5,4}x_1^2x_2^3+\alpha_{5,5}x_1x_2^4
 +\alpha_{5,6}x_2^5,\label{eq:poly5}\\
N_6={}&\alpha_{6,1}x_1^6+\alpha_{6,2}x_1^5x_2
 +\alpha_{6,3}x_1^4x_2^2+\alpha_{6,4}x_1^3x_2^3\nonumber\\
 &+\alpha_{6,5}x_1^2x_2^4+\alpha_{6,6}x_1x_2^5
 +\alpha_{6,7}x_2^6.\label{eq:poly6}
\end{align}

\subsection{Near-resonant approximation}\label{subsec:averaging}
We focus on the $2{:}1$ internal resonance, for which mode 1 responds near $\omega_D$ and mode 2 near $\omega_D/2$. Their displacements are written as
\begin{align}
x_1(t)&=\frac12\left(u_1e^{\ii\Om t}+u_1^*e^{-\ii\Om t}\right),\label{eq:ansatz1}\\
x_2(t)&=\frac12\left(u_2e^{\ii\Om t/2}+u_2^*e^{-\ii\Om t/2}\right),\label{eq:ansatz2}
\end{align}
where $u_1(t)$ and $u_2(t)$ vary slowly compared with the oscillations. We denote their amplitudes by $A=|u_1|$ and $B=|u_2|$. Under the slowly varying envelope approximation, $|\dot{u}_j|\ll\omega_D|u_j|$ and $\ddot{u}_j$ can be neglected compared with $\omega_D\dot{u}_j$.

We substitute these expressions into the nonlinear terms and retain contributions oscillating at $\omega_D$ in the first equation and at $\omega_D/2$ in the second. For example, the resonant parts of two quadratic and two cubic terms are
\begin{align}
x_2^2&\longrightarrow \frac14u_2^2e^{\ii\Om t},&
x_1x_2&\longrightarrow \frac14u_1u_2^*e^{\ii\Om t/2},\label{eq:quadraticproj}\\
x_1^3&\longrightarrow \frac38 A^2u_1e^{\ii\Om t},&
x_2^3&\longrightarrow \frac38 B^2u_2e^{\ii\Om t/2}.\label{eq:cubicproj}
\end{align}
The quadratic terms describe resonant coupling between the modes, while the cubic terms act as amplitude-dependent changes in their restoring forces. A similar calculation for the fourth, fifth, and sixth orders gives the retained terms summarized in Table~\ref{tab:nonlinear_coupling}. The full expansions are provided in Section S1 of the Supplementary Material.
\begin{table}
\centering
\caption{Carrier-retained representations of nonlinear monomials used in the compact two-mode reduction, with $A=|u_1|$ and $B=|u_2|$. Off-resonant terms and additional phase-sensitive resonant terms are not included in these replacements.}
\label{tab:nonlinear_coupling}
\begingroup\scriptsize
\setlength{\tabcolsep}{2.4pt}
\renewcommand{\arraystretch}{1.48}
\begin{tabularx}{\linewidth}{@{}*{5}{>{\centering\arraybackslash}X}@{}}
\toprule
\textbf{2nd order} & \textbf{3rd order} & \textbf{4th order} & \textbf{5th order} & \textbf{6th order}\\
\midrule
$x_1^2\simeq0$ & $x_1^3\simeq\tfrac34 A^2x_1$ & $x_1^4\simeq0$ & $x_1^5\simeq\tfrac58 A^4x_1$ & $x_1^6\simeq0$\\
$x_1x_2$ & $x_1^2x_2\simeq\tfrac12 A^2x_2$ & $x_1^3x_2\simeq\tfrac34 A^2x_1x_2$ & $x_1^4x_2\simeq\tfrac38 A^4x_2$ & $x_1^5x_2\simeq\tfrac58 A^4x_1x_2$\\
$x_2^2$ & $x_1x_2^2\simeq\tfrac12 B^2x_1$ & $x_1^2x_2^2\simeq\tfrac12 A^2x_2^2$ & $x_1^3x_2^2\simeq\tfrac38 A^2B^2x_1$ & $x_1^4x_2^2\simeq\tfrac38 A^4x_2^2$\\
& $x_2^3\simeq\tfrac34 B^2x_2$ & $x_1x_2^3\simeq\tfrac34 B^2x_1x_2$ & $x_1^2x_2^3\simeq\tfrac38 A^2B^2x_2$ & $x_1^3x_2^3\simeq\tfrac{9}{16}A^2B^2x_1x_2$\\
& & $x_2^4\simeq B^2x_2^2$ & $x_1x_2^4\simeq\tfrac38 B^4x_1$ & $x_1^2x_2^4\simeq\tfrac12 A^2B^2x_2^2$\\
& & & $x_2^5\simeq\tfrac58 B^4x_2$ & $x_1x_2^5\simeq\tfrac58 B^4x_1x_2$\\
& & & & $x_2^6\simeq\tfrac{15}{16}B^4x_2^2$\\
\bottomrule
\end{tabularx}
\endgroup
\end{table}

\subsection{Reduced equations}\label{subsec:reduced}
Using the terms retained above, the coupled equations can be written in the same form as a quadratically coupled system,
\begin{align}
 \ddot x_1+2\gamma_1\dot x_1+(\omega_1')^2x_1+\alpha_{22}'x_2^2&=F\cos(\Om t),\label{eq:short1}\\
 \ddot x_2+2\gamma_2\dot x_2+(\omega_2')^2x_2+\alpha_{12}'x_1x_2&=0.\label{eq:short2}
\end{align}

where the effective squared frequencies are
\begin{align}
 (\omega_1')^2={}&\omega_1^2+\frac34\alpha_{3,1}A^2+\frac12\alpha_{3,3}B^2
 +\frac58\alpha_{5,1}A^4\nonumber\\
 &+\frac38\alpha_{5,3}A^2B^2+\frac38\alpha_{5,5}B^4,\label{eq:omega1eff}\\
 (\omega_2')^2={}&\omega_2^2+\frac12\alpha_{3,2}A^2+\frac34\alpha_{3,4}B^2
 +\frac38\alpha_{5,2}A^4\nonumber\\
 &+\frac38\alpha_{5,4}A^2B^2+\frac58\alpha_{5,6}B^4.\label{eq:omega2eff}
\end{align}

and the effective quadratic coupling coefficients are
\begin{align}
 \alpha_{12}'={}&\alpha_{2,2}+\frac34\alpha_{4,2}A^2+\frac34\alpha_{4,4}B^2
 +\frac58\alpha_{6,2}A^4\nonumber\\
 &+\frac{9}{16}\alpha_{6,4}A^2B^2+\frac58\alpha_{6,6}B^4,\label{eq:coupling12}\\
 \alpha_{22}'={}&\alpha_{2,3}+\frac12\alpha_{4,3}A^2+\alpha_{4,5}B^2
 +\frac38\alpha_{6,3}A^4\nonumber\\
 &+\frac12\alpha_{6,5}A^2B^2+\frac{15}{16}\alpha_{6,7}B^4.\label{eq:coupling22}
\end{align}

Thus the cubic and quintic terms shift the modal frequencies, whereas the quartic and sixth-order terms change the coupling between the modes. These corrections depend on $A$ and $B$, so their importance changes as the response grows. Equations~\eqref{eq:short1}--\eqref{eq:coupling22} give the reduced model used in the numerical comparison.

\subsection{Existence condition for phononic frequency combs}\label{subsec:existence}
The reduced equations have the structure required for the quadratic-coupling existence condition derived by Qi \emph{et al.}~\cite{Qi2020}. We use that condition with the effective parameters defined above. Introducing $\tau=\gamma_1t$ and $\gamrat=\gamma_2/\gamma_1$, the detunings are
\begin{equation}\label{eq:detunings}
\Delta_1=\frac{\Om-\omega_1'}{\gamma_1},\qquad
\Delta_2=\frac{\Om/2-\omega_2'}{\gamma_1}
=\frac{\Delta_1}{2}+\kappa,\qquad
\kappa=\frac{\omega_1'-2\omega_2'}{2\gamma_1}.
\end{equation}
The normalized drive and response amplitudes are defined as
\begin{align}
 f_{\mathrm{an}}&=\frac{\alpha_{12}'}{8\gamma_1^2\Om^2}F,\label{eq:analyticforce}\\
 |\psi_1|&=\frac{\alpha_{12}'}{4\gamma_1\Om}A,\qquad
 |\psi_2|=\frac{\sqrt{\alpha_{12}'\alpha_{22}'}}{4\gamma_2\Om}B,\label{eq:psinorm}
\end{align}

The existence condition then becomes
\begin{equation}\label{eq:criterion}
 |\psi_2|^2\geq
 -\frac{\gamrat(1+\Delta_1^2)
 \bigl[1+\Delta_1^2+4\gamrat(1+\gamrat)\bigr]}
 {4(1+\gamrat)^2\bigl[1+\Delta_1^2+2\gamrat+2\Delta_1\Delta_2\bigr]}.
\end{equation}

For the right-hand side to be positive, the denominator factor must be negative. Substituting $\Delta_2=\Delta_1/2+\kappa$ gives a simple detuning condition,
\begin{equation}\label{eq:detuning_quadratic}
1+\Delta_1^2+2\gamrat+2\Delta_1\Delta_2
=2\Delta_1^2+2\kappa\Delta_1+(1+2\gamrat)\leq0.
\end{equation}
Solving this quadratic inequality gives
\begin{equation}\label{eq:detuningroots}
\frac{-\kappa-\sqrt{\kappa^2-2(1+2\gamrat)}}{2}
\leq\Delta_1\leq
\frac{-\kappa+\sqrt{\kappa^2-2(1+2\gamrat)}}{2},
\qquad \kappa^2\geq2(1+2\gamrat).
\end{equation}
The centre of this interval is $\Delta_{1,\mathrm{centre}}=-\kappa/2$, and its width is $\sqrt{\kappa^2-2(1+2\gamrat)}$. Converting to physical frequencies, and performing further analyses as shown in Supplementary Section S2, gives
\begin{align}
 \omega_{D,\mathrm{centre}}&=\frac{3\omega_1'+2\omega_2'}{4},\label{eq:center}\\
 R&=\left|\omega_2'-\frac{\omega_1'}{2}
 -\frac{\sqrt2\,\omega_1'}{\sqrt{Q_1Q_2}}\right|,\label{eq:range}\\
 \delta&=\frac{\omega_1'}{\sqrt{2Q_1Q_2}},\label{eq:delta}\\
 \omega_D^*&=\omega_1'+(\omega_1'-2\omega_2')
 \left(\frac{Q_2}{Q_1}\right)^2.\label{eq:criticaldrive}
\end{align}

The normalized subharmonic amplitude at the critical driving frequency is
\begin{equation}\label{eq:minpsi}
 |\psi_2|^*=\frac{\sqrt{\alpha_{12}'\alpha_{22}'}}{4\gamma_2\omega_D^*}\,B.
\end{equation}

The dependence on the higher-order terms becomes especially clear when the subharmonic amplitude dominates, $A\ll B$. If the effective coefficients are also expanded only up to $B^2$, they reduce to
\begin{equation}\label{eq:small_A_approx}
\begin{pmatrix}
(\omega_1')^2\\(\omega_2')^2\\\alpha'_{12}\\\alpha'_{22}
\end{pmatrix}
\simeq
\begin{pmatrix}
\omega_1^2+\tfrac12\alpha_{3,3}B^2\\
\omega_2^2+\tfrac34\alpha_{3,4}B^2\\
\alpha_{2,2}+\tfrac34\alpha_{4,4}B^2\\
\alpha_{2,3}+\alpha_{4,5}B^2
\end{pmatrix}.
\end{equation}

With the same small-amplitude approximation, the corresponding frequency and amplitude parameters are
\begin{align}
\omega_{D,\mathrm{centre}}
={}&\frac{3\omega_1+2\omega_2}{4}
+\left(\frac{3\alpha_{3,3}}{16\omega_1}
+\frac{3\alpha_{3,4}}{16\omega_2}\right)B^2,
\label{eq:center_B}\\
R={}&\left|\omega_2-\frac{\omega_1}{2}
-\frac{\sqrt2\,\omega_1}{\sqrt{Q_1Q_2}}
+\left(\frac{3\alpha_{3,4}}{8\omega_2}
-\frac{\alpha_{3,3}}{8\omega_1}
-\frac{\sqrt2\,\alpha_{3,3}}{4\omega_1\sqrt{Q_1Q_2}}\right)B^2\right|,
\label{eq:range_B}\\
\delta={}&\frac{\omega_1}{\sqrt{2Q_1Q_2}}
+\frac{\alpha_{3,3}B^2}{4\omega_1\sqrt{2Q_1Q_2}},
\label{eq:delta_B}\\
\omega_D^*={}&\omega_1+(\omega_1-2\omega_2)
\left(\frac{Q_2}{Q_1}\right)^2\nonumber\\
&+\left[\frac{\alpha_{3,3}}{4\omega_1}
+\left(\frac{\alpha_{3,3}}{4\omega_1}
-\frac{3\alpha_{3,4}}{4\omega_2}\right)
\left(\frac{Q_2}{Q_1}\right)^2\right]B^2,
\label{eq:drive_B}\\
|\psi_2|^*={}&\frac{\sqrt{\alpha_{2,2}\alpha_{2,3}}}
{4\gamma_2\omega_D^*}
\left[B+B^3\left(
\frac{3\alpha_{4,4}}{8\alpha_{2,2}}
+\frac{\alpha_{4,5}}{2\alpha_{2,3}}
\right)\right].
\label{eq:psi_B}
\end{align}

The minimum-amplitude driving frequency is obtained by minimizing the right-hand side of Eq.~\eqref{eq:criterion} with respect to $\Delta_1$, as detailed in Supplementary Section S2. Equations~\eqref{eq:center_B}--\eqref{eq:psi_B} apply to the weak-amplitude limit; the full effective coefficients in Eqs.~\eqref{eq:omega1eff}--\eqref{eq:coupling22} are used when higher powers of $A$ or $B$ are important. In this way the even- and odd-order corrections enter the existing phononic frequency-comb existence analysis.

\section{Numerical Simulation}
\label{sec:numerics}

The natural frequencies of the two coupled modes are estimated using the MoS$_2$ resonator model developed by Samanta et al. \cite{Samanta2018}. The squared natural frequency is given by
\begin{equation}\label{eq:mos2omega}
\begin{aligned}
\omega_0^2 ={}& \frac{E\varepsilon_r}{3\rho}\left(\frac{2\pi}{L}\right)^2
+\frac{EA_{\mathrm{DC}}^2}{6\rho}\left(\frac{2\pi}{L}\right)^4\\
&-\frac{5EA_{\mathrm{DC}}^4}{144\rho}\left(\frac{2\pi}{L}\right)^6
-\frac{C_2V_{\mathrm{DC}}^2}{\rho S},
\qquad C_2=\frac{\epsilon_0\mathcal A}{d^3L},
\end{aligned}
\end{equation}

where $\varepsilon_r$ is the residual strain, $\epsilon_0$ is the vacuum permittivity, $\mathcal A=LW$ is the membrane area, and $S=Wt$ is the cross-sectional area. The material and geometric parameters are given in Table~\ref{tab:parameters}. We use effective modal lengths $L_1=L$ and $L_2=1.5L$ and choose $V_{\rm DC}=13\,$V in the numerical benchmark. This gives $\omega_1/(2\pi)\simeq84.766\,$MHz and $\omega_2/(2\pi)\simeq43.760\,$MHz. These are illustrative mode frequencies obtained from Eq.~\eqref{eq:mos2omega}, not fitted measurements of the two-mode spectrum.
\begin{table}
\centering
\caption{Material/geometric inputs reported in the original MoS$_2$ example and numerical choices used in the present illustrative benchmark.}
\label{tab:parameters}
\small
\begin{tabularx}{0.8\linewidth}{@{}lXl@{}}
\toprule
Symbol & Quantity & Value\\
\midrule
$E$ & Young's modulus & $300\,\mathrm{GPa}$\\
$\rho$ & Density & $5000\,\mathrm{kg\,m^{-3}}$\\
$\varepsilon_r$ & Residual strain & $10^{-3}$\\
$\epsilon_0$ & Vacuum permittivity & $8.854\times10^{-12}\,\mathrm{F\,m^{-1}}$\\
$t$ & Membrane thickness & $1.3\,\mathrm{nm}$\\
$L$ & Reference length & $2\,\mu\mathrm{m}$\\
$W$ & Width & $L/4$\\
$d$ & Electrode gap & $185\,\mathrm{nm}$\\
$A_{\rm DC}$ & Static deflection & $6.64\times10^{-11}V_{\rm DC}^2\,\mathrm{m}$\\
$L_2/L_1$ & Effective length ratio (selected) & $1.5$\\
$V_{\rm DC}$ & Bias (selected) & $13\,\mathrm V$\\
\bottomrule
\end{tabularx}
\end{table}

For both the \emph{Long} model [Eqs.~\eqref{eq:full1}--\eqref{eq:poly6}] and the \emph{Short} model [Eqs.~\eqref{eq:short1}--\eqref{eq:coupling22}], we use cumulative nonlinear orders $N=3,4,5,6$ with identical initial-state construction. The benchmark uses $\gamma_2/\gamma_1=1$ and an inferred damping scale $\gamma_1\simeq9.613\times10^5\,\mathrm{s^{-1}}$ determined by the selected reference detunings $5$ and $-6.5$; it is not a measured damping constant. The quadratic coefficient scale is generated from the reference normalized frequency, while the cubic multiplier and higher-order strengths are set to
\begin{equation}\label{eq:etas}
\alpha=0.1,\qquad \eta_4=0.005,\qquad
\eta_5=0.1,\qquad \eta_6=0.001.
\end{equation}
The higher-order multipliers in Eq.~\eqref{eq:etas} are \emph{illustrative}. In particular, $\eta_4$ and $\eta_6$ differ from the earlier benchmark values $0.1$ and $0.01$; changing these values changes the nonlinear coefficients, rather than merely increasing solver precision. The scalar single-mode geometric coefficients reported for MoS$_2$ do not by themselves determine every intermodal $\alpha_{n,j}$, so the calculation is not a parameter-free simulation of the cited device.

With normalized time $\tau=\gamma_1 t$, the code integrates the Long and Short equations using fourth-order Runge--Kutta time stepping. Its force coordinate is defined by $F_{\rm}=2(\omega_D/\gamma_1)\,f_{\rm}$, where $f_{\rm}\in[0,40]$; it is not identical to the analytical normalization $f_{\rm an}$ in Eq.~\eqref{eq:analyticforce}. For the Short model the effective coefficients are evaluated from the single-carrier amplitude estimates
\begin{equation}\label{eq:amp_proxy}
\widehat A^2=x_1^2+\left(\frac{\dot x_1}{\omega_D}\right)^2,
\qquad
\widehat B^2=x_2^2+\left(\frac{2\dot x_2}{\omega_D}\right)^2.
\end{equation}
The fixed-drive contour calculations use $f_D=85.5\,$MHz, 41 force values, a total normalized duration $\tau=96$, and the last 60 units for the Hann-windowed FFT. The integration uses 48 steps and retains 16 samples per drive cycle; fourfold zero-padding is used only for frequency-axis interpolation. A separate onset scan uses $80$--$86\,$MHz (including $85.5\,$MHz), eleven coarse force values between 0 and 40, and local bisection refinement. That scan integrates to $\tau=80$ and tests late-time modulation of \emph{both} modal responses.

\section{Results and Discussion}
\label{sec:results}

\begin{figure}
    \centering
    \includegraphics[width=0.9\linewidth]{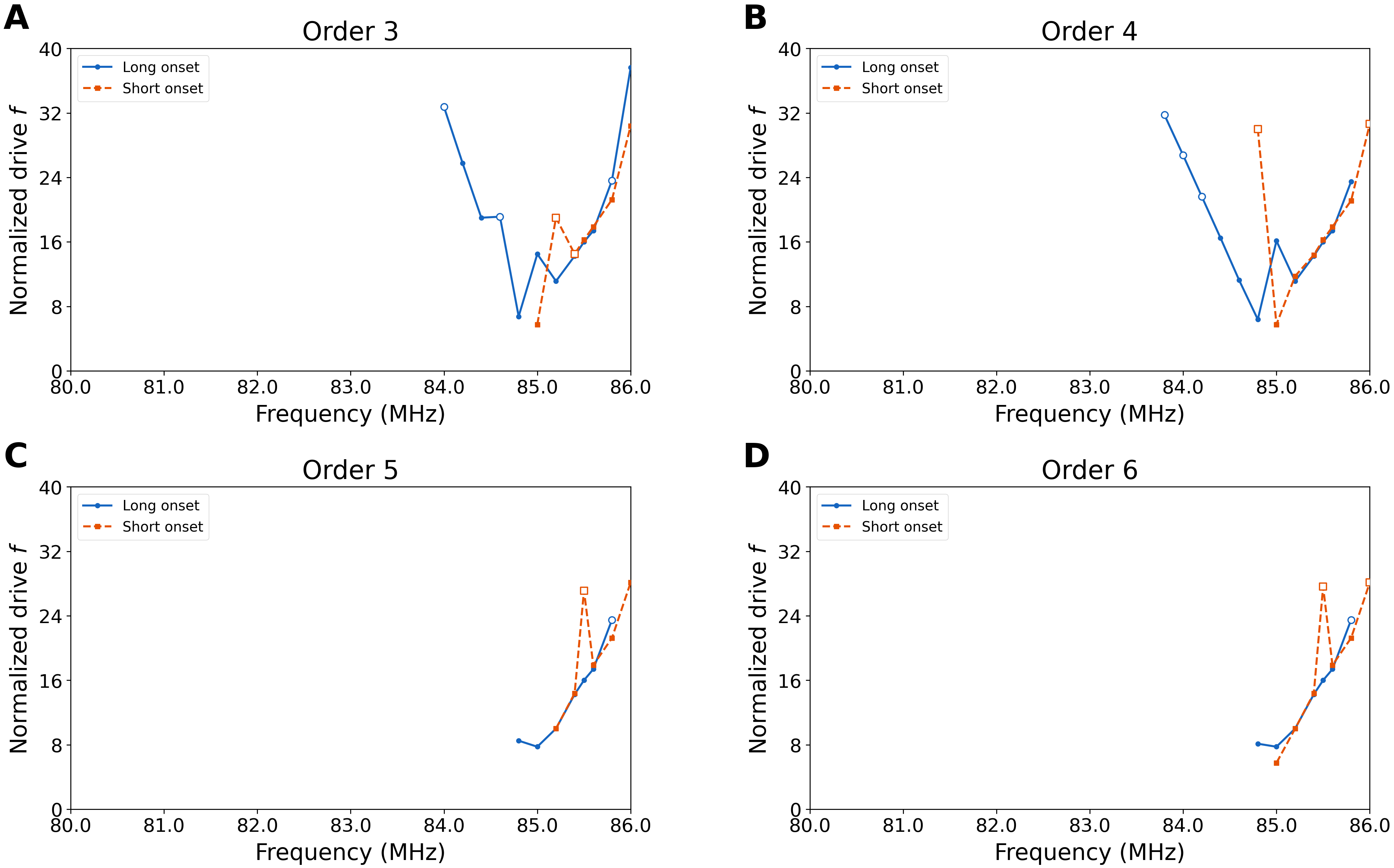}
    \caption{Numerical onset boundaries of phononic frequency combs for cumulative nonlinear orders (A) three, (B) four, (C) five, and (D) six. Solid blue and dashed orange curves represent the full nonlinear equations (Long) and the reduced equations (Short), respectively. The driving frequency spans $80$--$86\,\mathrm{MHz}$, while the normalized forcing varies from $0$ to $40$.}
    \label{fig1}
\end{figure}

\begin{figure}
    \centering
    \includegraphics[width=0.9\linewidth]{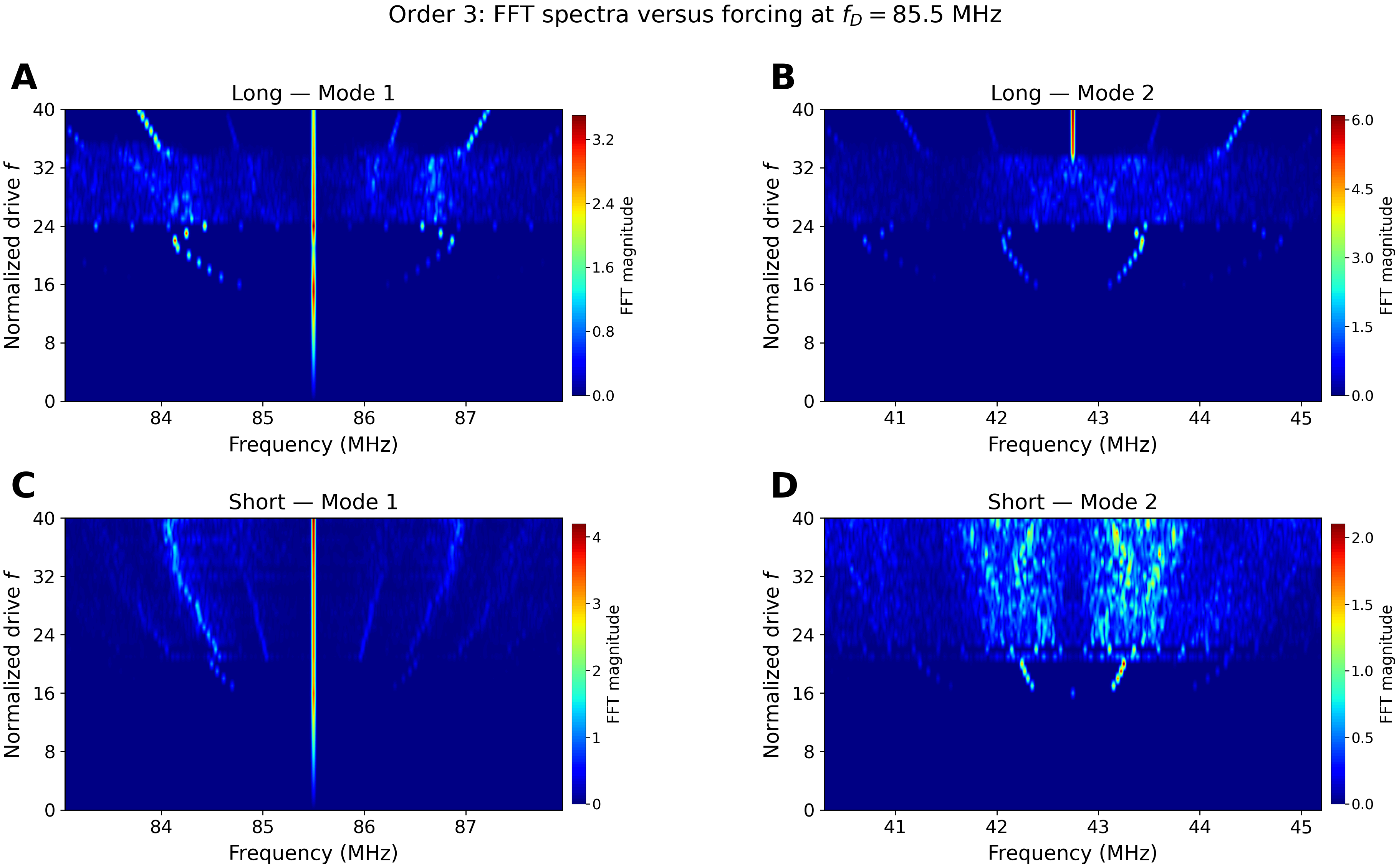}
    \caption{Spectra for the third-order nonlinear system at a fixed driving frequency of $f_D=85.5\,\mathrm{MHz}$, with normalized forcing $f\in[0,40]$. Panels (A) and (B) show the spectra of modes 1 and 2 obtained from the full nonlinear equations (Long), while panels (C) and (D) show the corresponding spectra from the reduced equations (Short). The frequency windows are centred near $85.5\,\mathrm{MHz}$ and $42.75\,\mathrm{MHz}$ for modes 1 and 2, respectively.}
    \label{fig2}
\end{figure}

\begin{figure}
    \centering
    \includegraphics[width=0.9\linewidth]{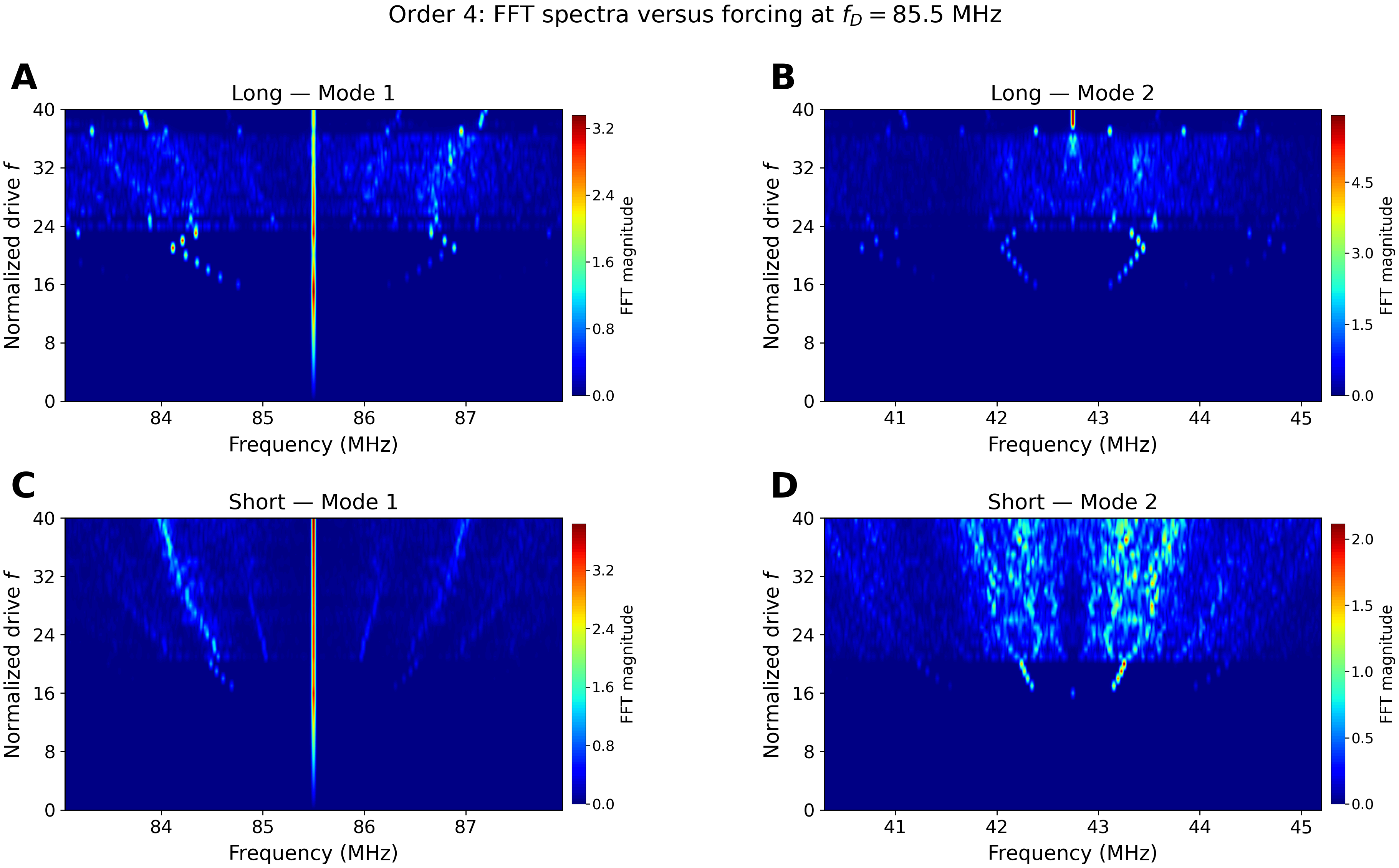}
    \caption{Spectra for the fourth-order nonlinear system at $f_D=85.5\,\mathrm{MHz}$ and normalized forcing $f\in[0,40]$. Panels (A,B) correspond to modes 1 and 2 of the full nonlinear equations (Long), while panels (C,D) represent the reduced equations (Short). The cumulative model includes nonlinear interactions through fourth order, with $\eta_4=0.005$.}
    \label{fig3}
\end{figure}

\begin{figure}
    \centering
    \includegraphics[width=0.9\linewidth]{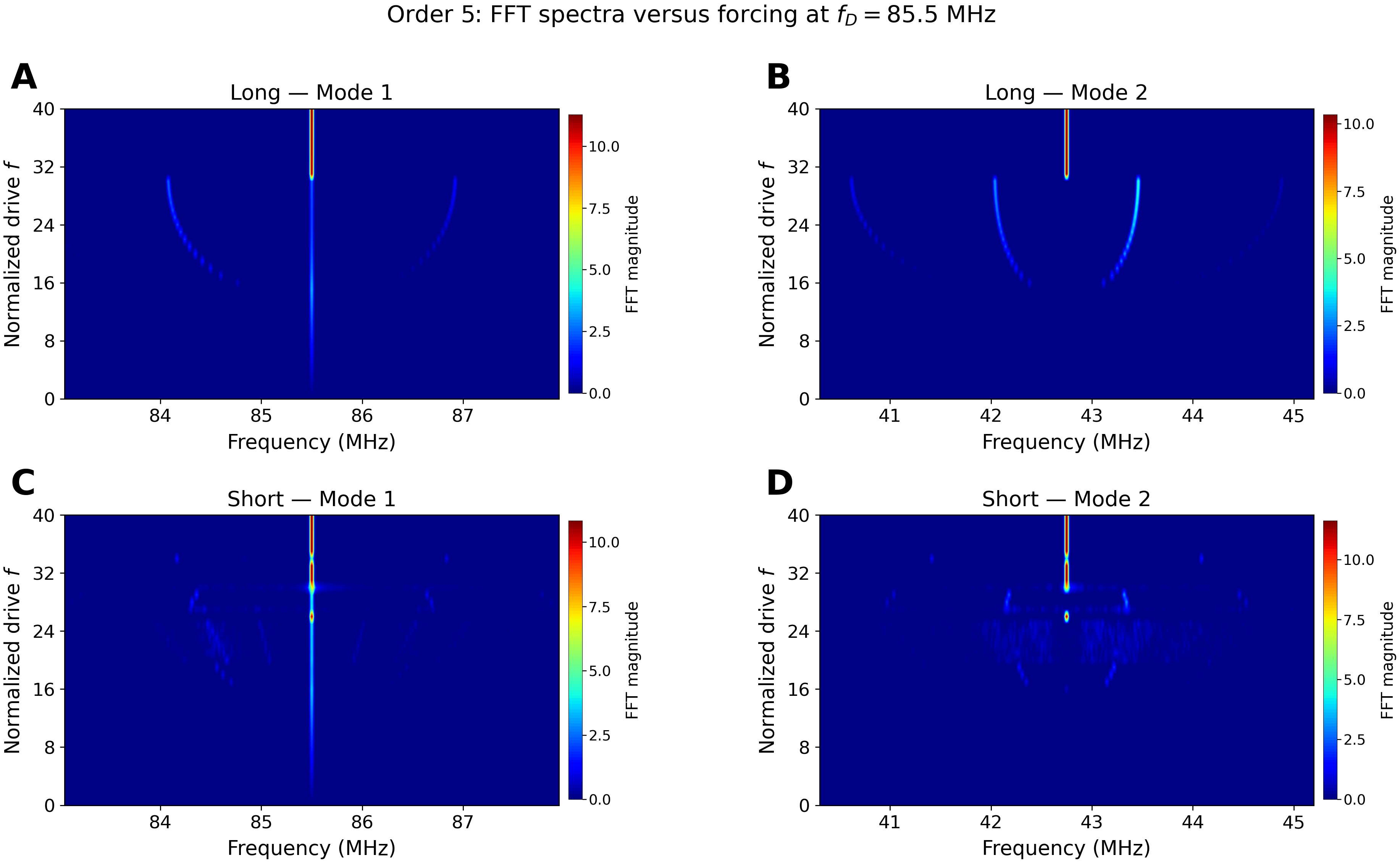}
    \caption{Spectra for the fifth-order nonlinear system at $f_D=85.5\,\mathrm{MHz}$ and normalized forcing $f\in[0,40]$. Panels (A,B) show modes 1 and 2 of the full nonlinear equations (Long), while panels (C,D) show the corresponding reduced equations (Short). Nonlinear interactions are retained through fifth order, with $\eta_4=0.005$ and $\eta_5=0.1$.}
    \label{fig4}
\end{figure}

\begin{figure}
    \centering
    \includegraphics[width=0.9\linewidth]{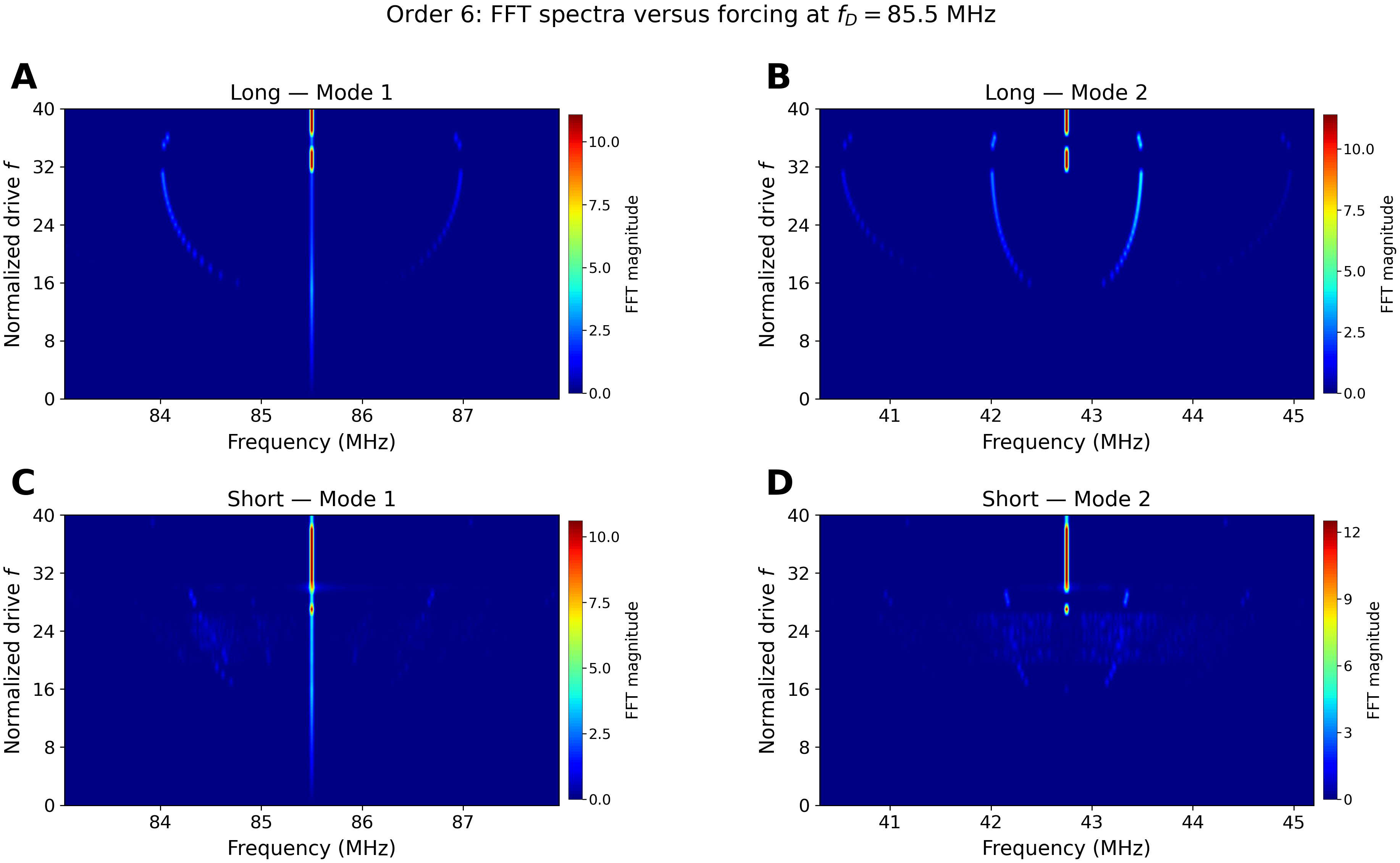}
    \caption{Spectra for the sixth-order nonlinear system at $f_D=85.5\,\mathrm{MHz}$ and normalized forcing $f\in[0,40]$. Panels (A,B) represent modes 1 and 2 of the full nonlinear equations (Long), while panels (C,D) show the corresponding reduced equations (Short). The cumulative model includes nonlinearities through sixth order, with $\eta_4=0.005$, $\eta_5=0.1$, and $\eta_6=0.001$.}
    \label{fig5}
\end{figure}

Figure~\ref{fig1} compares the numerical onset boundaries of the phononic frequency combs obtained from the full equation (long) and the reduced equation (short) for nonlinear orders of three to six. The onset is strongly dependent on the driving frequency, with the required forcing decreasing towards certain frequencies before increasing again. This behaviour reflects the importance of detuning in the nonlinear interaction between the two modes. Near the $2{:}1$ internal resonance, the transfer of energy from the directly driven mode to the subharmonic mode depends on the frequency mismatch, damping, and coupling strength. Consequently, the threshold does not vary monotonically with the driving frequency. For the third- and fourth-order systems, the Long and Short equations give comparable onset estimates over part of their common frequency range, indicating that the reduced model retains the principal features of the coupled-mode dynamics. At $f_D=85.5\,\mathrm{MHz}$, the Long equations give an onset near $f=16$ for all four orders, while the Short equations give approximately $f=16.25$ for orders three and four. The higher Short estimates for orders five and six remain sensitive to numerical refinement and cannot yet be regarded as converged thresholds.

The third-order system provides the starting point for understanding the development of phononic frequency combs, as shown in Fig.~\ref{fig2}. At lower forcing, the spectral response is concentrated near the driven and subharmonic frequencies. As the forcing increases, additional comb sidebands emerge around both frequency regions, forming approximately symmetric branches whose positions change with the driving amplitude. The simultaneous development of these features in both modes is consistent with the nonlinear energy exchange enabled by the quadratic coupling terms. The cubic nonlinearities further modify this interaction by introducing amplitude-dependent shifts in the effective modal frequencies. As a result, the spectral structure evolves with increasing forcing rather than remaining fixed around the initial resonances. Both the Long and Short equations reproduce the principal comb branches, although the Long model exhibits a broader distribution of spectral components at higher forcing, particularly in the subharmonic mode.

The effect of fourth-order nonlinearity is examined in Fig.~\ref{fig3}. The main comb branches remain similar to those of the third-order system, although changes are visible in their intensity and distribution over the forcing range. This relatively modest variation is consistent with the small fourth-order coefficient, $\eta_4=0.005$, used in the present calculations. According to the reduced model, the retained fourth-order nonlinearities modify the effective quadratic coupling coefficients through terms that depend on the modal amplitudes. Their influence therefore becomes more relevant as the oscillation amplitudes increase, altering the strength of interaction between the driven and subharmonic modes. The Long and Short equations produce comparable comb patterns at lower and intermediate forcing, while their differences become more apparent at higher forcing, particularly around the subharmonic frequency.

A more noticeable change occurs when fifth-order nonlinearities are included, as shown in Fig.~\ref{fig4}. Compared with the third- and fourth-order systems, the Long equations produce more localized comb sideband branches, while the broad spectral distribution previously observed at higher forcing is considerably reduced. The sidebands follow curved paths as the forcing changes, indicating that the frequencies associated with the nonlinear response depend on the oscillation amplitude. This behaviour is consistent with the quintic contributions to the effective modal frequencies derived in the mathematical model. Unlike even-order terms, which primarily modify the effective quadratic coupling within the adopted reduction, the retained fifth-order terms introduce additional amplitude-dependent stiffness corrections. These corrections change the effective detuning between the two modes and consequently influence the development and distribution of comb sidebands. The Short equations reproduce several corresponding spectral features, but noticeable differences remain at higher forcing, particularly in the subharmonic response.

Figure~\ref{fig5} presents the sixth-order system, where the comb spectra retain many of the features observed at fifth order. The Long equations continue to produce distinct curved sideband branches around both modal frequencies, while the reduced equations capture their principal frequency locations with some differences in spectral intensity and distribution. In the present reduction, the sixth-order nonlinearities contribute additional amplitude-dependent corrections to the effective quadratic coupling coefficients, complementing the frequency shifts introduced by the odd-order terms. The similarity between the fifth- and sixth-order spectra is consistent with the relatively small sixth-order coefficient, $\eta_6=0.001$, used in the numerical model. 

\section{Conclusions}
\label{sec:conclusions}

We have developed a reduced-order quadratic framework to examine the influence of higher-order nonlinearities on phononic frequency-comb generation in a driven two-mode system near a $2{:}1$ internal resonance. By retaining the dominant near-resonant interactions, we have shown that odd-order nonlinearities can be represented through amplitude-dependent corrections to the modal frequencies, whereas even-order nonlinearities modify the effective quadratic coupling coefficients. This reduction preserves the structure of the quadratic two-mode model while extending its applicability to polynomial nonlinear interactions of higher order.

The numerical results demonstrate that the generation of phononic combs depends on the combined effects of driving frequency, forcing amplitude, and nonlinear mode coupling. The onset boundaries identify the frequency-dependent forcing required to establish sustained coupled-mode modulation, while the Fourier spectra reveal how the comb sidebands develop and evolve as nonlinear terms through sixth order are included. In particular, the fifth- and sixth-order systems exhibit more localized spectral branches than the third- and fourth-order systems for the selected parameters, highlighting the influence of higher-order interactions on the distribution of comb frequencies. Comparisons between the full and reduced equations show good qualitative agreement in the principal comb features, supporting the effectiveness of the proposed approximation. However, differences in spectral intensities and some onset estimates indicate that its quantitative accuracy depends on the nonlinear order and operating conditions.

These findings provide a systematic way to incorporate higher-order nonlinear effects into the analysis of phononic frequency-comb existence conditions without retaining every nonlinear term in the reduced equations. The framework offers a useful basis for studying the formation of combs in mechanical resonators with complex nonlinear responses. Further work involving experimentally determined coupling coefficients, detailed numerical convergence tests, and phase-resolved spectral analysis would help to establish their quantitative accuracy for specific devices.

\bibliographystyle{unsrt}
\bibliography{references}

\appendix

\clearpage

\begin{center}

{\LARGE\bfseries
Supplementary Material:
Reduced-Order Quadratic Representation of
Higher-Order Nonlinear Dynamical Systems
for Phononic Frequency Comb Generation
\par}

\vspace{0.15cm}

{\large
Madhurendra Mishra$^{1}$ and
Adarsh Ganesan$^{2}$
\par}

\vspace{0.1cm}

{\normalsize
$^{1}$Department of Physics,
Sri Guru Tegh Bahadur Khalsa College,
University of Delhi, India

$^{2}$Department of Electrical and
Electronics Engineering,
BITS Pilani, Dubai Campus, UAE
\par}

\vspace{0.15cm}

{\normalsize
\textit{Corresponding author:}
adarsh@dubai.bits-pilani.ac.in
\par}

\end{center}

\vspace{0.3cm}

\section*{Supplementary Section S1: Expressions for the nonlinear terms in terms of \(x_{1}\), \(x_{2}\) and \(x_{1}x_{2}\)}

The equations of motion corresponding to two coupled modes can be written as

\begin{equation}\label{eq:S1-1}\tag{S1-1}
{\ddot{x}}_{1} + 2\gamma_{1}{\dot{x}}_{1} + \omega_{1}^{2}x_{1} + N.I. = Fcos\left( \omega_{D}t \right)
\end{equation}

\begin{equation}\label{eq:S1-2}\tag{S1-2}
{\ddot{x}}_{2} + 2\gamma_{2}{\dot{x}}_{2} + \omega_{2}^{2}x_{2} + N.I. = 0
\end{equation}

Here,
\[
\begin{aligned}
\mathbf{N.I.}={}&
\alpha_{2,1}x_1^2+\alpha_{2,2}x_1x_2+\alpha_{2,3}x_2^2+\\
&\alpha_{3,1}x_1^3+\alpha_{3,2}x_1^2x_2+\alpha_{3,3}x_1x_2^2+\alpha_{3,4}x_2^3+\\
&\alpha_{4,1}x_1^4+\alpha_{4,2}x_1^3x_2+\alpha_{4,3}x_1^2x_2^2\\
&\quad+\alpha_{4,4}x_1x_2^3+\alpha_{4,5}x_2^4+\\
&\alpha_{5,1}x_1^5+\alpha_{5,2}x_1^4x_2+\alpha_{5,3}x_1^3x_2^2\\
&\quad+\alpha_{5,4}x_1^2x_2^3+\alpha_{5,5}x_1x_2^4+\alpha_{5,6}x_2^5+\\
&\alpha_{6,1}x_1^6+\alpha_{6,2}x_1^5x_2+\alpha_{6,3}x_1^4x_2^2\\
&\quad+\alpha_{6,4}x_1^3x_2^3+\alpha_{6,5}x_1^2x_2^4\\
&\quad+\alpha_{6,6}x_1x_2^5+\alpha_{6,7}x_2^6+\\
&\ldots
\end{aligned}
\]

The modes are written as

\begin{equation}\label{eq:S1-3}\tag{S1-3}
x_{1} = \frac{1}{2}\left( u_{1}e^{i\omega_{D}t} + u_{1}^{*}e^{- i\omega_{D}t} \right)
\end{equation}

\begin{equation}\label{eq:S1-4}\tag{S1-4}
x_{2} = \frac{1}{2}\left( u_{2}e^{i\frac{\omega_{D}}{2}t} + u_{2}^{*}e^{- i\frac{\omega_{D}}{2}t} \right)
\end{equation}

The derivatives \({\dot{\mathbf{x}}}_{\mathbf{1}}\), \({\dot{\mathbf{x}}}_{\mathbf{2}}\), \({\ddot{\mathbf{x}}}_{\mathbf{1}}\) and \({\ddot{\mathbf{x}}}_{\mathbf{2}}\) are obtained as

\begin{equation}\label{eq:S1-5}\tag{S1-5}
\begin{aligned}
{\dot{x}}_{1} &= \frac{1}{2}\left({\renewcommand{\arraystretch}{0.85}\begin{matrix*}[l]
i\omega_{D}u_{1}e^{i\omega_{D}t} + {\dot{u}}_{1}e^{i\omega_{D}t} - i\omega_{D}u_{1}e^{- i\omega_{D}t} \\
+ {\dot{u}}_{1}^{*}e^{- i\omega_{D}t}\end{matrix*}}\right)
\end{aligned}
\end{equation}

\begin{equation}\label{eq:S1-6}\tag{S1-6}
\begin{aligned}
{\dot{x}}_{2} &= \frac{1}{2}\left({\renewcommand{\arraystretch}{0.85}\begin{matrix*}[l]
i\frac{\omega_{D}}{2}u_{2}e^{i\frac{\omega_{D}}{2}t} + {\dot{u}}_{2}e^{i\frac{\omega_{D}}{2}t} \\

- i\frac{\omega_{D}}{2}u_{2}e^{- i\frac{\omega_{D}}{2}t} + {\dot{u}}_{2}^{*}e^{- i\frac{\omega_{D}}{2}t}\end{matrix*}}\right)
\end{aligned}
\end{equation}

\begin{equation}\label{eq:S1-7}\tag{S1-7}
\begin{aligned}
{\ddot{x}}_{1} &= \frac{1}{2}\left({\renewcommand{\arraystretch}{0.85}\begin{matrix*}[l]
{\ddot{u}}_{1}e^{i\omega_{D}t} + 2i\omega_{D}{\dot{u}}_{1}e^{i\omega_{D}t} - \omega_{D}^{2}u_{1}e^{i\omega_{D}t} \\
+ {\ddot{u}}_{1}^{*}e^{- i\omega_{D}t} - 2i\omega_{D}{\dot{u}}_{1}^{*}e^{- i\omega_{D}t} \\

- \omega_{D}^{2}u_{1}^{*}e^{- i\omega_{D}t}\end{matrix*}}\right)
\end{aligned}
\end{equation}

\begin{equation}\label{eq:S1-8}\tag{S1-8}
\begin{aligned}
{\ddot{x}}_{2} &= \frac{1}{2}\left({\renewcommand{\arraystretch}{0.85}\begin{matrix*}[l]
{\ddot{u}}_{2}e^{i\frac{\omega_{D}}{2}t} + i\omega_{D}{\dot{u}}_{2}e^{i\frac{\omega_{D}}{2}t} \\

- \frac{\omega_{D}^{2}}{4}u_{2}e^{i\frac{\omega_{D}}{2}t} + {\ddot{u}}_{2}^{*}e^{- i\frac{\omega_{D}}{2}t} \\

- i\omega_{D}{\dot{u}}_{2}^{*}e^{- i\frac{\omega_{D}}{2}t} - \frac{\omega_{D}^{2}}{4}u_{2}^{*}e^{- i\frac{\omega_{D}}{2}t}\end{matrix*}}\right)
\end{aligned}
\end{equation}

We collect only the terms containing \(e^{i\omega_{D}t}\) and \(e^{i\frac{\omega_{D}}{2}t}\).\label{we-collect-only-the-terms-containing-eiomega_dt-and-eifracomega_d2t.}

\begin{equation}\label{eq:S1-9}\tag{S1-9}
{\dot{x}}_{1} = \frac{1}{2}\left( i\omega_{D}u_{1}e^{i\omega_{D}t} + {\dot{u}}_{1}e^{i\omega_{D}t} \right)
\end{equation}

\begin{equation}\label{eq:S1-10}\tag{S1-10}
{\dot{x}}_{2} = \frac{1}{2}\left( i\frac{\omega_{D}}{2}u_{2}e^{i\frac{\omega_{D}}{2}t} + {\dot{u}}_{2}e^{i\frac{\omega_{D}}{2}t} \right)
\end{equation}

\begin{equation}\label{eq:S1-11}\tag{S1-11}
{\ddot{x}}_{1} = \frac{1}{2}\left( {\ddot{u}}_{1}e^{i\omega_{D}t} + 2i\omega_{D}{\dot{u}}_{1}e^{i\omega_{D}t} - \omega_{D}^{2}u_{1}e^{i\omega_{D}t} \right)
\end{equation}

\begin{equation}\label{eq:S1-12}\tag{S1-12}
\begin{aligned}
{\ddot{x}}_{2} &= \frac{1}{2}\left({\renewcommand{\arraystretch}{0.85}\begin{matrix*}[l]
{\ddot{u}}_{2}e^{i\frac{\omega_{D}}{2}t} + i\omega_{D}{\dot{u}}_{2}e^{i\frac{\omega_{D}}{2}t} \\

- \frac{\omega_{D}^{2}}{4}u_{2}e^{i\frac{\omega_{D}}{2}t}\end{matrix*}}\right)
\end{aligned}
\end{equation}

Similarly, the coupling terms can also be estimated as follows.\label{similarly-the-coupling-terms-can-also-be-estimated-as-follows.}

\smallskip\noindent\textbf{2\textsuperscript{nd} Order Nonlinearity}\par\nobreak

\begin{equation}\label{eq:S1-13}\tag{S1-13}
x_{1}^{2} = \frac{1}{4}\left( u_{1}^{2}e^{2i\omega_{D}t} + {u_{1}^{*}}^{2}e^{- 2i\omega_{D}t} + 2\left| u_{1} \right|^{2} \right)\sim 0
\end{equation}

\begin{equation}\label{eq:S1-14}\tag{S1-14}
\begin{aligned}
x_{1}x_{2} &= \frac{1}{4}\left({\renewcommand{\arraystretch}{0.85}\begin{matrix*}[l]
u_{1}^{*}u_{2}^{*}e^{- 3i\frac{\omega_{D}}{2}t} + u_{1}^{*}u_{2}e^{- i\frac{\omega_{D}}{2}t} \\
+ u_{1}u_{2}^{*}e^{i\frac{\omega_{D}}{2}t} + u_{1}u_{2}e^{3i\frac{\omega_{D}}{2}t}\end{matrix*}}\right)
\\[1pt]
&\sim \frac{1}{4}u_{1}u_{2}^{*}e^{i\frac{\omega_{D}}{2}t}
\end{aligned}
\end{equation}

\begin{equation}\label{eq:S1-15}\tag{S1-15}
\begin{aligned}
x_{2}^{2} &= \frac{1}{4}\left( u_{2}^{2}e^{i\omega_{D}t} + {u_{2}^{*}}^{2}e^{- i\omega_{D}t} + 2\left| u_{2} \right|^{2} \right)
\\[1pt]
&\sim \frac{1}{4}u_{2}^{2}e^{i\omega_{D}t}
\end{aligned}
\end{equation}

\smallskip\noindent\textbf{3\textsuperscript{rd} Order Nonlinearity}\par\nobreak

\begin{equation}\label{eq:S1-16}\tag{S1-16}
\begin{aligned}
x_{1}^{3} &= \frac{1}{8}\left({\renewcommand{\arraystretch}{0.85}\begin{matrix*}[l]
u_{1}^{3}e^{3i\omega_{D}t} + 3\left| u_{1} \right|^{2}u_{1}e^{i\omega_{D}t} \\
+ 3\left| u_{1} \right|^{2}u_{1}^{*}e^{- i\omega_{D}t} + {u_{1}^{*}}^{3}e^{- 3i\omega_{D}t}\end{matrix*}}\right)
\\[1pt]
&\sim \frac{3}{8}\left| u_{1} \right|^{2}u_{1}e^{i\omega_{D}t}
\end{aligned}
\end{equation}

\begin{equation}\label{eq:S1-17}\tag{S1-17}
\begin{aligned}
x_{1}^{2}x_{2} &= \frac{1}{8}\left({\renewcommand{\arraystretch}{0.85}\begin{matrix*}[l]
u_{1}^{2}u_{2}e^{i\frac{5\omega_{D}}{2}t} + {u_{1}^{*}}^{2}u_{2}e^{- i\frac{3\omega_{D}}{2}t} \\
+ 2\left| u_{1} \right|^{2}u_{2}e^{i\frac{\omega_{D}}{2}t} + u_{1}^{2}u_{2}^{*}e^{i\frac{3\omega_{D}}{2}t} \\
+ {u_{1}^{*}}^{2}u_{2}^{*}e^{- i\frac{5\omega_{D}}{2}t} + 2\left| u_{1} \right|^{2}u_{2}^{*}e^{- i\frac{\omega_{D}}{2}t}\end{matrix*}}\right)
\\[1pt]
&\sim \frac{1}{4}\left| u_{1} \right|^{2}u_{2}e^{i\frac{\omega_{D}}{2}t}
\end{aligned}
\end{equation}

\begin{equation}\label{eq:S1-18}\tag{S1-18}
\begin{aligned}
x_{1}x_{2}^{2} &= \frac{1}{8}\left({\renewcommand{\arraystretch}{0.85}\begin{matrix*}[l]
u_{1}u_{2}^{2}e^{2i\omega_{D}t} + u_{1}^{*}u_{2}^{2} + u_{1}{u_{2}^{*}}^{2} \\
+ u_{1}^{*}{u_{2}^{*}}^{2}e^{- 2i\omega_{D}t} + 2\left| u_{2} \right|^{2}u_{1}e^{i\omega_{D}t} \\
+ 2\left| u_{2} \right|^{2}u_{1}^{*}e^{- i\omega_{D}t}\end{matrix*}}\right)
\\[1pt]
&\sim \frac{1}{4}\left| u_{2} \right|^{2}u_{1}e^{i\omega_{D}t}
\end{aligned}
\end{equation}

\begin{equation}\label{eq:S1-19}\tag{S1-19}
\begin{aligned}
x_{2}^{3} &= \frac{1}{8}\left({\renewcommand{\arraystretch}{0.85}\begin{matrix*}[l]
u_{2}^{3}e^{i\frac{3\omega_{D}}{2}t} + 3\left| u_{2} \right|^{2}u_{2}e^{i\frac{\omega_{D}}{2}t} \\
+ 3\left| u_{2} \right|^{2}u_{2}^{*}e^{- i\frac{\omega_{D}}{2}t} + {u_{2}^{*}}^{3}e^{- i\frac{3\omega_{D}}{2}t}\end{matrix*}}\right)
\\[1pt]
&\sim \frac{3}{8}\left| u_{2} \right|^{2}u_{2}e^{i\frac{\omega_{D}}{2}t}
\end{aligned}
\end{equation}

\smallskip\noindent\textbf{4\textsuperscript{th} Order Nonlinearity}\par\nobreak

\begin{equation}\label{eq:S1-20}\tag{S1-20}
\begin{aligned}
x_{1}^{4} &= \frac{1}{16}\left({\renewcommand{\arraystretch}{0.85}\begin{matrix*}[l]
u_{1}^{4}e^{4i\omega_{D}t} + 4\left| u_{1} \right|^{2}u_{1}^{2}e^{2i\omega_{D}t} + 6\left| u_{1} \right|^{4} \\
+ 4\left| u_{1} \right|^{2}{u_{1}^{*}}^{2}e^{- 2i\omega_{D}t} + {u_{1}^{*}}^{4}e^{- 4i\omega_{D}t}\end{matrix*}}\right)
\\[1pt]
&\sim 0
\end{aligned}
\end{equation}

\begin{equation}\label{eq:S1-21}\tag{S1-21}
\begin{aligned}
x_{1}^{3}x_{2} &= \frac{1}{16}\left({\renewcommand{\arraystretch}{0.85}\begin{matrix*}[l]
u_{1}^{3}e^{3i\omega_{D}t} + 3\left| u_{1} \right|^{2}u_{1}e^{i\omega_{D}t} \\
+ 3\left| u_{1} \right|^{2}u_{1}^{*}e^{- i\omega_{D}t} + {u_{1}^{*}}^{3}e^{- 3i\omega_{D}t}\end{matrix*}}\right)
\\[1pt]
&\qquad \left( u_{2}e^{i\frac{\omega_{D}}{2}t} + u_{2}^{*}e^{- i\frac{\omega_{D}}{2}t} \right)
\\[1pt]
&\sim \frac{3}{16}\left| u_{1} \right|^{2}\left( u_{1}u_{2}^{*} \right)e^{i\frac{\omega_{D}}{2}t}
\end{aligned}
\end{equation}

\begin{equation}\label{eq:S1-22}\tag{S1-22}
\begin{aligned}
x_{1}^{2}x_{2}^{2} &= \frac{1}{16}\left({\renewcommand{\arraystretch}{0.85}\begin{matrix*}[l]
u_{1}^{*}u_{2}^{*}e^{- 3i\frac{\omega_{D}}{2}t} + u_{1}^{*}u_{2}e^{- i\frac{\omega_{D}}{2}t} \\
+ u_{1}u_{2}^{*}e^{i\frac{\omega_{D}}{2}t} + u_{1}u_{2}e^{3i\frac{\omega_{D}}{2}t}\end{matrix*}}\right)
\\[1pt]
&\qquad \left({\renewcommand{\arraystretch}{0.85}\begin{matrix*}[l]
u_{1}^{*}u_{2}^{*}e^{- 3i\frac{\omega_{D}}{2}t} + u_{1}^{*}u_{2}e^{- i\frac{\omega_{D}}{2}t} \\
+ u_{1}u_{2}^{*}e^{i\frac{\omega_{D}}{2}t} + u_{1}u_{2}e^{3i\frac{\omega_{D}}{2}t}\end{matrix*}}\right)
\\[1pt]
&\sim \frac{1}{8}\left| u_{1} \right|^{2}u_{2}^{2}e^{i\omega_{D}t} + \frac{1}{16}\left( u_{1}^{2}{u_{2}^{*}}^{2} \right)e^{i\omega_{D}t}
\end{aligned}
\end{equation}

\begin{equation}\label{eq:S1-23}\tag{S1-23}
\begin{aligned}
x_{1}x_{2}^{3} &= \frac{1}{16}\left( u_{1}e^{i\omega_{D}t} + u_{1}^{*}e^{- i\omega_{D}t} \right)
\\[1pt]
&\qquad \left({\renewcommand{\arraystretch}{0.85}\begin{matrix*}[l]
u_{2}^{3}e^{\frac{3i\omega_{D}}{2}t} + 3\left| u_{2} \right|^{2}u_{2}e^{\frac{i\omega_{D}}{2}t} \\
+ 3\left| u_{2} \right|^{2}u_{2}^{*}e^{- \frac{i\omega_{D}}{2}t} + {u_{2}^{*}}^{3}e^{- \frac{3i\omega_{D}}{2}t}\end{matrix*}}\right)
\\[1pt]
&\sim \frac{1}{16}\left( u_{1}^{*}u_{2}^{3} \right)e^{i\frac{\omega_{D}}{2}t} + \frac{3}{16}\left| u_{2} \right|^{2}\left( u_{1}u_{2}^{*} \right)e^{i\frac{\omega_{D}}{2}t}
\end{aligned}
\end{equation}

\begin{equation}\label{eq:S1-24}\tag{S1-24}
\begin{aligned}
x_{2}^{4} &= \frac{1}{16}\left({\renewcommand{\arraystretch}{0.85}\begin{matrix*}[l]
u_{2}^{4}e^{2i\omega_{D}t} + 4\left| u_{2} \right|^{2}u_{2}^{2}e^{i\omega_{D}t} + 6\left| u_{2} \right|^{4} \\
+ 4\left| u_{2} \right|^{2}{u_{2}^{*}}^{2}e^{- i\omega_{D}t} + {u_{1}^{*}}^{4}e^{- 2i\omega_{D}t}\end{matrix*}}\right)
\\[1pt]
&\sim \frac{1}{4}\left| u_{2} \right|^{2}u_{2}^{2}e^{i\omega_{D}t}
\end{aligned}
\end{equation}

\smallskip\noindent\textbf{5\textsuperscript{th} Order Nonlinearity}\par\nobreak

\begin{equation}\label{eq:S1-25}\tag{S1-25}
\begin{aligned}
x_{1}^{5} &= \frac{1}{32}\left({\renewcommand{\arraystretch}{0.85}\begin{matrix*}[l]
u_{1}^{5}e^{5i\omega_{D}t} + 5u_{1}^{4}u_{1}^{*}e^{3i\omega_{D}t} \\
+ 10u_{1}^{3}{u_{1}^{*}}^{2}e^{i\omega_{D}t} + 10u_{1}^{2}{u_{1}^{*}}^{3}e^{- i\omega_{D}t} \\
+ \ {u_{1}^{*}}^{5}e^{- 5\omega_{D}t}\end{matrix*}}\right)
\\[1pt]
&\sim \frac{5}{16}\left| u_{1} \right|^{4}u_{1}e^{i\omega_{D}t}
\end{aligned}
\end{equation}

\begin{equation}\label{eq:S1-26}\tag{S1-26}
\begin{aligned}
x_{1}^{4}x_{2} &= \frac{1}{32}\left({\renewcommand{\arraystretch}{0.85}\begin{matrix*}[l]
u_{1}^{4}e^{4i\omega_{D}t} + 4\left| u_{1} \right|^{2}u_{1}^{2}e^{2i\omega_{D}t} + 6\left| u_{1} \right|^{4} \\
+ 4\left| u_{1} \right|^{2}{u_{1}^{*}}^{2}e^{- 2i\omega_{D}t} + {u_{1}^{*}}^{4}e^{- 4i\omega_{D}t}\end{matrix*}}\right)
\\[1pt]
&\qquad \left( u_{2}e^{i\frac{\omega_{D}}{2}t} + u_{2}^{*}e^{- i\frac{\omega_{D}}{2}t} \right)
\\[1pt]
&\sim \frac{3}{16}\left| u_{1} \right|^{4}u_{2}e^{i\frac{\omega_{D}}{2}t}
\end{aligned}
\end{equation}

\begin{equation}\label{eq:S1-27}\tag{S1-27}
\begin{aligned}
x_{1}^{3}x_{2}^{2} &= \frac{1}{32}\left({\renewcommand{\arraystretch}{0.85}\begin{matrix*}[l]
u_{1}^{3}e^{3i\omega_{D}t} + 3\left| u_{1} \right|^{2}u_{1}e^{i\omega_{D}t} \\
+ 3\left| u_{1} \right|^{2}u_{1}^{*}e^{- i\omega_{D}t} + {u_{1}^{*}}^{3}e^{- 3i\omega_{D}t}\end{matrix*}}\right)
\\[1pt]
&\qquad \left( u_{2}^{2}e^{i\omega_{D}t} + {u_{2}^{*}}^{2}e^{- i\omega_{D}t} + 2\left| u_{2} \right|^{2} \right)
\\[1pt]
&\sim \frac{3}{16}\left| u_{1} \right|^{2}\left| u_{2} \right|^{2}u_{1}e^{i\omega_{D}t}
\end{aligned}
\end{equation}

\begin{equation}\label{eq:S1-28}\tag{S1-28}
\begin{aligned}
x_{1}^{2}x_{2}^{3} &= \frac{1}{32}\left({\renewcommand{\arraystretch}{0.85}\begin{matrix*}[l]
u_{2}^{3}e^{i\frac{3\omega_{D}}{2}t} + 3\left| u_{2} \right|^{2}u_{2}e^{i\frac{\omega_{D}}{2}t} \\
+ 3\left| u_{2} \right|^{2}u_{2}^{*}e^{- i\frac{\omega_{D}}{2}t} + {u_{2}^{*}}^{3}e^{- i\frac{3\omega_{D}}{2}t}\end{matrix*}}\right)
\\[1pt]
&\qquad \left( u_{1}^{2}e^{2i\omega_{D}t} + {u_{1}^{*}}^{2}e^{- 2i\omega_{D}t} + 2\left| u_{1} \right|^{2} \right)
\\[1pt]
&\sim \frac{1}{32}u_{1}^{2}{u_{2}^{*}}^{3}e^{i\frac{\omega_{D}}{2}t} + \frac{3}{16}\left| u_{1} \right|^{2}\left| u_{2} \right|^{2}u_{2}e^{i\frac{\omega_{D}}{2}t}
\end{aligned}
\end{equation}

\begin{equation}\label{eq:S1-29}\tag{S1-29}
\begin{aligned}
x_{1}x_{2}^{4} &= \frac{1}{32}\left({\renewcommand{\arraystretch}{0.85}\begin{matrix*}[l]
u_{2}^{4}e^{2i\omega_{D}t} + 4\left| u_{2} \right|^{2}u_{2}^{2}e^{i\omega_{D}t} + 6\left| u_{2} \right|^{4} \\
+ 4\left| u_{2} \right|^{2}{u_{2}^{*}}^{2}e^{- i\omega_{D}t} + {u_{1}^{*}}^{4}e^{- 2i\omega_{D}t}\end{matrix*}}\right)
\\[1pt]
&\qquad \left( u_{1}e^{i\omega_{D}t} + u_{1}^{*}e^{- i\omega_{D}t} \right)
\\[1pt]
&\sim \frac{1}{32}u_{2}^{4}u_{1}^{*}e^{i\omega_{D}t} + \frac{3}{16}\left| u_{2} \right|^{4}u_{1}e^{i\omega_{D}t}
\end{aligned}
\end{equation}

\begin{equation}\label{eq:S1-30}\tag{S1-30}
\begin{aligned}
x_{2}^{5} &= \frac{1}{32}\left({\renewcommand{\arraystretch}{0.85}\begin{matrix*}[l]
u_{2}^{5}e^{\frac{5i\omega_{D}}{2}t} + 5u_{2}^{4}u_{2}^{*}e^{\frac{3i\omega_{D}}{2}t} \\
+ 10u_{2}^{3}{u_{2}^{*}}^{2}e^{\frac{i\omega_{D}}{2}t} \\
+ 10u_{2}^{2}{u_{2}^{*}}^{3}e^{- \frac{i\omega_{D}}{2}t} + \ {u_{2}^{*}}^{5}e^{- \frac{5\omega_{D}}{2}t}\end{matrix*}}\right)
\\[1pt]
&\sim \frac{5}{16}\left| u_{2} \right|^{4}u_{2}e^{\frac{i\omega_{D}}{2}t}
\end{aligned}
\end{equation}

\smallskip\noindent\textbf{6\textsuperscript{th} Order Nonlinearity}\par\nobreak

\begin{equation}\label{eq:S1-31}\tag{S1-31}
\begin{aligned}
x_{1}^{6} &= \frac{1}{64}\left({\renewcommand{\arraystretch}{0.85}\begin{matrix*}[l]
u_{1}^{6}e^{6i\omega_{D}t} + 6u_{1}^{5}u_{1}^{*}e^{4i\omega_{D}t} \\
+ 15u_{1}^{4}{u_{1}^{*}}^{2}e^{2i\omega_{D}t} + 20u_{1}^{3}{u_{1}^{*}}^{3} \\
+ 15u_{1}^{2}{u_{1}^{*}}^{4}e^{- 2i\omega_{D}t} + 6u_{1}{u_{1}^{*}}^{5}e^{- 4i\omega_{D}t} \\
+ \ {u_{1}^{*}}^{6}e^{- 6\omega_{D}t}\end{matrix*}}\right)
\\[1pt]
&\sim 0
\end{aligned}
\end{equation}

\begin{equation}\label{eq:S1-32}\tag{S1-32}
\begin{aligned}
x_{1}^{5}x_{2} &= \frac{1}{64}\left({\renewcommand{\arraystretch}{0.85}\begin{matrix*}[l]
u_{1}^{5}e^{5i\omega_{D}t} + 5u_{1}^{4}u_{1}^{*}e^{3i\omega_{D}t} \\
+ 10u_{1}^{3}{u_{1}^{*}}^{2}e^{i\omega_{D}t} + 10u_{1}^{2}{u_{1}^{*}}^{3}e^{- i\omega_{D}t} \\
+ \ {u_{1}^{*}}^{5}e^{- 5\omega_{D}t}\end{matrix*}}\right)
\\[1pt]
&\qquad \left( u_{2}e^{i\frac{\omega_{D}}{2}t} + u_{2}^{*}e^{- i\frac{\omega_{D}}{2}t} \right)
\\[1pt]
&\sim \frac{5}{32}\left| u_{1} \right|^{4}\left( u_{1}u_{2}^{*} \right)e^{i\frac{\omega_{D}}{2}t}
\end{aligned}
\end{equation}

\begin{equation}\label{eq:S1-33}\tag{S1-33}
\begin{aligned}
x_{1}^{4}x_{2}^{2} &= \frac{1}{64}\left({\renewcommand{\arraystretch}{0.85}\begin{matrix*}[l]
u_{1}^{4}e^{4i\omega_{D}t} + 4\left| u_{1} \right|^{2}u_{1}^{2}e^{2i\omega_{D}t} + 6\left| u_{1} \right|^{4} \\
+ 4\left| u_{1} \right|^{2}{u_{1}^{*}}^{2}e^{- 2i\omega_{D}t} + {u_{1}^{*}}^{4}e^{- 4i\omega_{D}t}\end{matrix*}}\right)
\\[1pt]
&\qquad \left( u_{2}^{2}e^{i\omega_{D}t} + {u_{2}^{*}}^{2}e^{- i\omega_{D}t} + 2\left| u_{2} \right|^{2} \right)
\\[1pt]
&\sim \frac{1}{64}\left( 6\left| u_{1} \right|^{4}u_{2}^{2}e^{i\omega_{D}t} + 4\left| u_{1} \right|^{2}u_{1}^{2}{u_{2}^{*}}^{2}e^{i\omega_{D}t} \right)
\end{aligned}
\end{equation}

\begin{equation}\label{eq:S1-34}\tag{S1-34}
\begin{aligned}
x_{1}^{3}x_{2}^{3} &= \frac{1}{64}\left({\renewcommand{\arraystretch}{0.85}\begin{matrix*}[l]
u_{1}^{3}e^{3i\omega_{D}t} + 3\left| u_{1} \right|^{2}u_{1}e^{i\omega_{D}t} \\
+ 3\left| u_{1} \right|^{2}u_{1}^{*}e^{- i\omega_{D}t} + {u_{1}^{*}}^{3}e^{- 3i\omega_{D}t}\end{matrix*}}\right)
\\[1pt]
&\qquad \left({\renewcommand{\arraystretch}{0.85}\begin{matrix*}[l]
u_{2}^{3}e^{i\frac{3\omega_{D}}{2}t} + 3\left| u_{2} \right|^{2}u_{2}e^{i\frac{\omega_{D}}{2}t} \\
+ 3\left| u_{2} \right|^{2}u_{2}^{*}e^{- i\frac{\omega_{D}}{2}t} + {u_{2}^{*}}^{3}e^{- i\frac{3\omega_{D}}{2}t}\end{matrix*}}\right)
\\[1pt]
&\sim \frac{1}{64}\left( 3\left| u_{1} \right|^{2}u_{1}^{*}u_{2}^{3}e^{i\frac{\omega_{D}}{2}t} + 9\left| u_{1} \right|^{2}\left| u_{2} \right|^{2}\left( u_{1}u_{2}^{*} \right)e^{i\frac{\omega_{D}}{2}t} \right)
\end{aligned}
\end{equation}

\begin{equation}\label{eq:S1-35}\tag{S1-35}
\begin{aligned}
x_{1}^{2}x_{2}^{4} &= \frac{1}{64}\left({\renewcommand{\arraystretch}{0.85}\begin{matrix*}[l]
u_{2}^{4}e^{2i\omega_{D}t} + 4\left| u_{2} \right|^{2}u_{2}^{2}e^{i\omega_{D}t} + 6\left| u_{2} \right|^{4} \\
+ 4\left| u_{2} \right|^{2}{u_{2}^{*}}^{2}e^{- i\omega_{D}t} + {u_{1}^{*}}^{4}e^{- 2i\omega_{D}t}\end{matrix*}}\right)
\\[1pt]
&\qquad \left( u_{1}^{2}e^{2i\omega_{D}t} + {u_{1}^{*}}^{2}e^{- 2i\omega_{D}t} + 2\left| u_{1} \right|^{2} \right)
\\[1pt]
&\sim \frac{1}{64}\left( 4\left| u_{2} \right|^{2}{u_{2}^{*}}^{2}u_{1}^{2} + 8\left| u_{1} \right|^{2}\left| u_{2} \right|^{2}u_{2}^{2}e^{i\omega_{D}t} \right)
\end{aligned}
\end{equation}

\begin{equation}\label{eq:S1-36}\tag{S1-36}
\begin{aligned}
x_{1}x_{2}^{5} &= \frac{1}{64}\left({\renewcommand{\arraystretch}{0.85}\begin{matrix*}[l]
u_{2}^{5}e^{\frac{5i\omega_{D}}{2}t} + 5u_{2}^{4}u_{2}^{*}e^{\frac{3i\omega_{D}}{2}t} \\
+ 10u_{2}^{3}{u_{2}^{*}}^{2}e^{\frac{i\omega_{D}}{2}t} \\
+ 10u_{2}^{2}{u_{2}^{*}}^{3}e^{- \frac{i\omega_{D}}{2}t} + \ {u_{2}^{*}}^{5}e^{- \frac{5\omega_{D}}{2}t}\end{matrix*}}\right)
\\[1pt]
&\qquad \left( u_{1}e^{i\omega_{D}t} + u_{1}^{*}e^{- i\omega_{D}t} \right)
\\[1pt]
&\sim \frac{5}{32}\left| u_{2} \right|^{4}\left( u_{1}u_{2}^{*} \right)e^{\frac{i\omega_{D}}{2}t}
\end{aligned}
\end{equation}

\begin{equation}\label{eq:S1-37}\tag{S1-37}
\begin{aligned}
x_{2}^{6} &= \frac{1}{64}\left({\renewcommand{\arraystretch}{0.85}\begin{matrix*}[l]
u_{2}^{6}e^{3i\omega_{D}t} + 6u_{2}^{5}u_{2}^{*}e^{2i\omega_{D}t} \\
+ 15u_{2}^{4}{u_{2}^{*}}^{2}e^{i\omega_{D}t} + 20u_{2}^{3}{u_{2}^{*}}^{3} \\
+ 15u_{2}^{2}{u_{2}^{*}}^{4}e^{- i\omega_{D}t} + 6u_{2}{u_{2}^{*}}^{5}e^{- 2i\omega_{D}t} \\
+ \ {u_{2}^{*}}^{6}e^{- 3\omega_{D}t}\end{matrix*}}\right)
\\[1pt]
&\sim \frac{15}{64}\left| u_{2} \right|^{4}u_{2}^{2}e^{i\omega_{D}t}
\end{aligned}
\end{equation}

\section*{Supplementary Section S2: Frequency Parameters of the Existence Boundary of Phononic Frequency Combs}

\textbf{S2-1. Center frequency of existence band of phononic frequency combs}

The frequency matching condition for phononic frequency combs is governed by eq. (S2.1).

\begin{equation}\label{eq:S2-1}\tag{S2.1}
\left( 1 + \Delta_{1}^{2} + 2\gamma + 2\Delta_{1}\Delta_{2} \right) \leq 0
\end{equation}

Since \(\Delta_{2} = \frac{\Delta_{1}}{2} + \overline{\kappa}\), we have the following expression.

\begin{equation}\label{eq:S2-2}\tag{S2.2}
\left( 1 + \Delta_{1}^{2} + 2\gamma + 2\Delta_{1}\left( \frac{{\overline{\Delta}}_{1}}{2} + \overline{\kappa} \right) \right) \leq 0
\end{equation}

\begin{equation}\label{eq:S2-3}\tag{S2.3}
\left( 2\Delta_{1}^{2} + 2\Delta_{1}\overline{\kappa} + (2\gamma + 1) \right) \leq 0
\end{equation}

\begin{equation}\label{eq:S2-4}\tag{S2.4}
\frac{- \overline{\kappa} - \sqrt{{\overline{\kappa}}^{2} - 2(2\gamma + 1)}}{2} \leq \Delta_{1} \leq \frac{- \overline{\kappa} + \sqrt{{\overline{\kappa}}^{2} - 2(2\gamma + 1)}}{2}
\end{equation}

The center frequency of existence band of phononic frequency combs is calculated as follows.

\begin{equation}\label{eq:S2-5}\tag{S2.5}
\Delta_{1,center} = - \frac{\overline{\kappa}}{2}
\end{equation}

\[\Rightarrow \frac{\omega_{D,center} - \omega_{1}}{\gamma_{1}} = - \frac{\omega_{1} - 2\omega_{2}}{4\gamma_{1}}\]

\[\Rightarrow \omega_{D,center} = \frac{{3\omega}_{1} + 2\omega_{2}}{4}\]

\textbf{S2-2. Range of existence band of phononic frequency combs}

The range of existence band of phononic frequency combs is calculated as follows.

\begin{equation}\label{eq:S2-6}\tag{S2.6}
\Delta_{1,max} - \Delta_{1,min} = \sqrt{{\overline{\kappa}}^{2} - 2(2\gamma + 1)}
\end{equation}

\[\Rightarrow \left( \omega_{D,max} - \omega_{D,min} \right)^{2} = \gamma_{1}^{2}\left( \frac{\left( \omega_{1} - 2\omega_{2} \right)^{2}}{4\gamma_{1}^{2}} - \frac{2\left( 2\gamma_{2} + \gamma_{1} \right)}{\gamma_{1}} \right)\]

\[\Rightarrow R^{2} = \left( \frac{\left( \omega_{1} - 2\omega_{2} \right)^{2} - 8\left( 2\gamma_{1}\gamma_{2} + \gamma_{1}^{2} \right)}{4} \right)\]

\[\Rightarrow R^{2} = \left( \frac{\omega_{1}^{2} - 4\omega_{1}\omega_{2} + 4\omega_{2}^{2} - 8\left( \frac{2\omega_{1}\omega_{2}}{Q_{1}Q_{2}} + \frac{\omega_{1}^{2}}{4Q_{1}^{2}} \right)}{4} \right)\]

\[\Rightarrow R^{2} = \left( \omega_{2}^{2} - \left( 1 + \frac{4}{Q_{1}Q_{2}} \right)\omega_{1}\omega_{2} + \frac{\omega_{1}^{2}}{4}\left( 1 - \frac{2}{Q_{1}^{2}} \right) \right)\]

\[\Rightarrow R^{2} = \left( \omega_{2}^{2} - \omega_{1}\omega_{2} + \frac{\omega_{1}^{2}}{4} \right) - \left( \frac{4\omega_{1}\omega_{2}}{Q_{1}Q_{2}} \right) - \left( \frac{\omega_{1}^{2}}{2Q_{1}^{2}} \right)\]

For \(Q_{1}^{2} \gg 2\), \(\frac{\omega_{1}^{2}}{4} \gg \frac{\omega_{1}^{2}}{2Q_{1}^{2}}\). Hence, \(\frac{\omega_{1}^{2}}{2Q_{1}^{2}}\) can be neglected from eq. (S2.6).

\begin{equation}\label{eq:S2-7}\tag{S2.7}
R^{2} = \left( \omega_{2}^{2} - \omega_{1}\omega_{2} + \frac{\omega_{1}^{2}}{4} \right) - \left( \frac{4\omega_{1}\omega_{2}}{Q_{1}Q_{2}} \right)
\end{equation}

\[\Rightarrow R^{2} = \left( \omega_{2}^{2} - \omega_{1}\omega_{2}\left( 1 + \frac{4}{Q_{1}Q_{2}} \right) + \frac{\omega_{1}^{2}}{4} \right)\]

The quadratic on \(\omega_{2}\) in the RHS of eq. (S2.7) can be written as

\begin{equation}\label{eq:S2-8}\tag{S2.8}
\begin{gathered}
R^{2} = \left( \omega_{2} - \omega_{1}\left[ \frac{\left( 1 + \frac{4}{Q_{1}Q_{2}} \right) + \sqrt{\left( 1 + \frac{4}{Q_{1}Q_{2}} \right)^{2} - 1}}{2} \right] \right)^{2} \\ R^{2} = \left( \omega_{2} - \frac{\omega_{1}}{2}\left( \left( 1 + \frac{4}{Q_{1}Q_{2}} \right) + \sqrt{\frac{8}{Q_{1}Q_{2}} + \frac{16}{Q_{1}^{2}Q_{2}^{2}}} \right) \right)^{2}
\end{gathered}
\end{equation}

For \(Q_{1}Q_{2} \gg 2\), \(\frac{8}{Q_{1}Q_{2}} \gg \frac{16}{Q_{1}^{2}Q_{2}^{2}}\). Hence, the term \(\frac{16}{Q_{1}^{2}Q_{2}^{2}}\) can be neglected in eq. (S2.8).

\begin{equation}\label{eq:S2-9}\tag{S2.9}
R^{2} = \left( \omega_{2} - \frac{\omega_{1}}{2}\left( 1 + \frac{4}{Q_{1}Q_{2}} + \sqrt{\frac{8}{Q_{1}Q_{2}}} \right) \right)^{2}
\end{equation}

For \(Q_{1}Q_{2} \gg 2\), \(\sqrt{\frac{8}{Q_{1}Q_{2}}} \gg \frac{4}{Q_{1}Q_{2}}\). Hence, the term \(\frac{4}{Q_{1}Q_{2}}\) can be neglected in eq. (S2.9).

\begin{equation}\label{eq:S2-10}\tag{S2.10}
R^{2} = \left( \omega_{2} - \frac{\omega_{1}}{2}\left( 1 + \sqrt{\frac{8}{Q_{1}Q_{2}}} \right) \right)^{2}
\end{equation}

An expression for range can hence be obtained as follows.

\begin{equation}\label{eq:S2-11}\tag{S2.11}
R = \left| \omega_{2} - \frac{\omega_{1}}{2} - \frac{\sqrt{2}\omega_{1}}{\sqrt{Q_{1}Q_{2}}} \right|
\end{equation}

\textbf{S2-3. Critical detuning for the generation of phononic frequency combs}

For \(\omega_{2} - \frac{\omega_{1}}{2} < 0\), the right bound of regime of phononic frequency combs is closer to \(\omega_{1}\), i.e., \(\omega_{D,center} + \frac{R}{2}\).

\begin{equation}\label{eq:S2-12}\tag{S2.12}
\begin{gathered}
\omega_{D,edge} = \omega_{D,center} + \frac{R}{2} = \frac{{3\omega}_{1} + 2\omega_{2}}{4} - \frac{\omega_{2}}{2} + \frac{\omega_{1}}{4} + \frac{\omega_{1}}{\sqrt{2Q_{1}Q_{2}}} \\ \Rightarrow \omega_{D,edge} = \omega_{D,center} + \frac{R}{2} = \omega_{1} + \frac{\omega_{1}}{\sqrt{2Q_{1}Q_{2}}}
\end{gathered}
\end{equation}

For \(\omega_{2} - \frac{\omega_{1}}{2} > 0\), the left bound of regime of phononic frequency combs is closer to \(\omega_{1}\), i.e., \(\omega_{D,center} - \frac{R}{2}\).

\begin{equation}\label{eq:S2-13}\tag{S2.13}
\begin{gathered}
\omega_{D,edge} = \omega_{D,center} - \frac{R}{2} = \frac{{3\omega}_{1} + 2\omega_{2}}{4} - \frac{\omega_{2}}{2} + \frac{\omega_{1}}{4} + \frac{\omega_{1}}{\sqrt{2Q_{1}Q_{2}}} \\ \Rightarrow \omega_{D,edge} = \omega_{D,center} - \frac{R}{2} = \omega_{1} + \frac{\omega_{1}}{\sqrt{2Q_{1}Q_{2}}}
\end{gathered}
\end{equation}

Hence, the minimum detuning from \(\omega_{1}\) that is required for exciting phononic combs can be obtained as

\begin{equation}\label{eq:S2-14}\tag{S2.14}
\delta = \left| \omega_{D,edge} - \omega_{1} \right| = \frac{\omega_{1}}{\sqrt{2Q_{1}Q_{2}}}
\end{equation}

\subsection*{S2-4. Critical drive frequency \(\mathbf{\omega}_{\mathbf{D}}^{\mathbf{*}}\) at which \(\left| \mathbf{\psi}_{\mathbf{2}} \right|\) reaches its minimum value}

\begin{equation}\label{eq:S2-15}\tag{S2-15}
\left| \psi_{2} \right|^{2} \geq - \frac{\gamma_{21}\left( 1 + \Delta_{1}^{2} \right)\left( 1 + \Delta_{1}^{2} + 4\gamma_{21}\left( 1 + \gamma_{21} \right) \right)}{4\left( 1 + \gamma_{21} \right)^{2}\left( 1 + \Delta_{1}^{2} + 2\gamma_{21} + 2\Delta_{1}\Delta_{2} \right)}
\end{equation}

\begin{equation}\label{eq:S2-16}\tag{S2-16}
\left| \psi_{2} \right|^{2} \geq - \frac{\gamma_{21}}{4\left( 1 + \gamma_{21} \right)^{2}}\frac{\left( \Delta_{1}^{4} + \Delta_{1}^{2}\left( 4\gamma_{21}^{2} + 4\gamma_{21} + 2 \right) + \left( 4\gamma_{21}^{2} + 4\gamma_{21} + 1 \right) \right)}{\left( \Delta_{1}^{2}(2) + \Delta_{1}(2\kappa) + (2\gamma_{21} + 1) \right)}
\end{equation}

\[\frac{{d\left| \psi_{2} \right|}^{2}}{{d\Delta}_{1}} = 0\]

\begin{equation}\label{eq:S2-17}\tag{S2-17}
\begin{aligned}
& \left[ \Delta_{1}^{2}(2) + \Delta_{1}(2\kappa) + \left( 2\gamma_{21} + 1 \right) \right]\left[ \Delta_{1}^{3}(4) + \Delta_{1}\left( 8\gamma_{21}^{2} + 8\gamma_{21} + 4 \right) \right]\\[4pt]
&- \left[ \Delta_{1}^{4} + \Delta_{1}^{2}\left( 4\gamma_{21}^{2} + 4\gamma_{21} + 2 \right) + \left( 4\gamma_{21}^{2} + 4\gamma_{21} + 1 \right) \right]\left[ \Delta_{1}(4) + 2\kappa \right] = 0
\end{aligned}
\end{equation}

\begin{equation}\label{eq:S2-18}\tag{S2-18}
\begin{aligned}
& \Delta_{1}^{5}(4) + \Delta_{1}^{4}(6\kappa) + \Delta_{1}^{3}\left( 8\gamma_{21} + 4 \right) + \Delta_{1}^{2}\left( 8\gamma_{21}^{2}\kappa + 8\gamma_{21}\kappa + 4\kappa \right) \\
& + \Delta_{1}\left( 16\gamma_{21}^{3} + 8\gamma_{21}^{2} \right) - \left( 8\gamma_{21}^{2}\kappa + 8\gamma_{21}\kappa + 2\kappa \right) = 0
\end{aligned}
\end{equation}

For \(Q_{2} \ll Q_{1}\), we have \(\gamma_{21} \gg \kappa,\Delta_{1}\). Hence, we have

\begin{equation}\label{eq:S2-19}\tag{S2-19}
\Delta_{1}^{2}(\kappa) + \Delta_{1}\left( 2\gamma_{21} \right) - (\kappa) = 0
\end{equation}

Hence, by solving this quadratic equation, we have \(\Delta_{1} = \frac{1}{2}\frac{\kappa}{\gamma_{21}^{2}}\ \)and \(\Delta_{1} = - 2\left( \frac{\gamma_{21}}{\kappa} \right)\). Since \(- 2\left( \frac{\gamma_{21}}{\kappa} \right)\) is very large and not feasible, we can only consider \(\Delta_{1} = \frac{1}{2}\frac{\kappa}{\gamma_{21}^{2}}\). \(\omega_{D}\) can thus be found as \(\omega_{D} = \omega_{1} + \frac{1}{4}(\omega_{1} - 2\omega_{2})\left( \frac{\omega_{1}}{\omega_{2}} \right)^{2}\left( \frac{Q_{2}}{Q_{1}} \right)^{2}\). Since \(\frac{\omega_{1}}{\omega_{2}}\sim 2\), we have

\begin{equation}\label{eq:S2-20}\tag{S2-20}
\omega_{D} = \omega_{1} + (\omega_{1} - 2\omega_{2})\left( \frac{Q_{2}}{Q_{1}} \right)^{2}
\end{equation}

\end{document}